\documentclass[preprint,aps]{revtex4}
\usepackage{epsfig}
\begin{document}
\title{Possible Force-Entropy Correlation}
\author{Enrique Canessa\footnote{E-mail: canessae@ictp.trieste.it}}
\affiliation{The Abdus Salam International Centre for Theoretical Physics,
Trieste, Italy
\vspace{2cm}
}

\begin{abstract}
A statistical thermodynamic approach of moving particles forming an elastic 
body is presented which leads to reveal molecular-mechanical properties 
of classical and nonextensive dynamical systems. We derive the Boltzmann-Gibbs
(BG) entropy form and relate it to Newton's law of motion in relation to a
distinct tensile force acting on the systems at constant volume and number
of particles.  Tsallis generalization of the BG entropy is deduced assuming
the thermal energy of the particles to be proportional to their energy states
by the nonextensivity factor $q-1$. 
\end{abstract}
\maketitle

\section{Overview}
It was pointed out long ago by Einstein that the foundations of statistical 
mechanics should lie on the underlying dynamics \cite{Coh02}.  All statistical 
mechanics results should derive from the dynamics if the systems considered
are systems consisting of moving particles.

Within the classical Boltzmann-Gibbs (BG) statistics, it is still unkown 
how to derive from say Newton's law of motion the remarkable expression 
for the thermodynamic entropy \cite{Tsa03} 
\begin{equation}\label{eq:bg}
S_{BG}/k_{B} = - \sum_{i=1}^{W} p_{i} \; \ln \; p_{i}  = 
   \left< \ln \; \frac{1}{p_{i}} \right> \;\;\; ,
\end{equation}
where mean values $< (\cdots ) >$ correspond to $\sum_{i} (\cdots ) \; p_{i}$.
Even if this functional form has been profusely verified on isolated systems
in thermal equilibrium, a rigorous derivation on how
$S_{BG}$ descends from dynamics has been limited to a dilute gas only
\cite{Coh02}. It is usually approximated via the statistics of a large
ensemble of identical replicas (see, {\it e.g.}, \cite{Man88}).
Only recently, there has been computational evidence on the fact that BG thermal
equilibrium distribution descends from Newton law \cite{Bal2004}. 

Also unclear at present \cite{Coh02,Tsa03}, is the connection between the
dynamics of a particle system and the Tsallis nonextensive generalization
of the BG entropy
\begin{equation}\label{eq:tsallis}
S_{q}/k_{B}=\left< \ln_{q} \; \frac{1}{p_{i}} \right> = 
            \sum_{i=1}^{W} p_{i} \; \ln_{q} \; \frac{1}{p_{i}} = 
            \frac{1 - \sum_{i=1}^{W} p_{i}^{q}}{q-1} \;\;\; , 
\end{equation}
where 
\begin{equation}\label{eq:lnq}
\ln_{q} \; p_{i} \equiv \frac{p_{i}^{1-q} - 1}{1-q} \;\;\; ,
\end{equation}
which for $q = 1$, it coincides 
with $S_{BG}$ giving positive Lyapunov exponents \cite{Tsa88}.  The
nonextensivity is given by the pseudo-additivity property
$S_{q}(A+B)=S_{q}(A)+S_{q}(B)+\frac{1-q}{k_{B}}S_{q}(A)S_{q}(B)$;
$A$ and $B$ being two independent systems.

In order to shed some light on this physics dilemma,
we propose a simple statistical thermodynamic model
where moving particles of known type are assumed to form an 
elastic specimen (or body).  It is shown that this type of approach leads
to reveal intrinsic molecular-mechanical properties on classical and 
nonextensive dynamical systems in relation to a distinct tensile force
acting on these systems at constant volume and number of particles.  
In particular we derive the entropy of Eq.(\ref{eq:bg}) and relate it 
to Newton's second law $\vec{F} =m \vec{a}$.  
A new scenario for the entropic index $q$ in Tsallis 
statistics in terms of the energy of the system is proposed.

\section{Formalism}

Let us consider a system of non-interacting moving particles and 
let $p_{i}$ be the probability that the system
is in the microstate $i (= 1, \cdots , W)$ given by the product of 
two positive functions $u_{i}$ and $w_{i}$ and satisfying the
normalization condition
\begin{equation}\label{eq:probability}
\sum_{i=1}^{W}p_{i}=\sum_{i=1}^{W}u_{i}w_{i} = 1 \;\;\; .
\end{equation}
Derivative of this relation with respect to changes in particle
displacements $x$, considering particles of mass $m$ to move at 
velocity $v=\frac{\partial x}{\partial t}$ and acceleration 
$a=\frac{\partial v}{\partial t}$, gives 
\begin{equation}\label{eq:sumderp}
\sum_{i=1}^{W}\frac{\partial p_{i}}{\partial x} = 
   \sum_{i=1}^{W}p_{i} \; \frac{\partial}{\partial x} \ln \; p_{i} = 0  
\end{equation}
or, alternatively,
\begin{equation}
\sum_{i=1}^{W} (\frac{\partial u_{i}}{\partial x}) \; \frac{p_{i}}{u_{i}} = 
- \sum_{i=1}^{W} (\frac{\partial w_{i}}{\partial x}) \; \frac{p_{i}}{w_{i}} 
 \;\;\; .
\end{equation}
Summing up both sides by 
$\sum_{i=1} (\frac{\partial p_{i}}{\partial x}) \; \ln \; u_{i}$ 
it is straightforward to show that
\begin{equation}\label{eq:means}
\frac{\partial }{\partial x} < \ln \; u_{i} > =
- \; \frac{\partial }{\partial x} < \ln \; w_{i} >  + 
\sum_{i}^{W} (\frac{\partial p_{i}}{\partial x}) \; \ln \; p_{i} \;\;\; .
\end{equation}

Let us also consider the {\it purely} thermodynamic relation for
the Helmholtz free energy $A=E-TS$, assuming the temperature $T$ and
volume $V$ to be independent variables with $E$ the internal energy,
$A$ the Helmholtz free energy and $S$ the entropy of the system.  
Therefore from Eq.(\ref{eq:means}) we can readily identify (or, alternatively, 
define) three quantities 
\begin{eqnarray}
A/k_{B}T & \rightarrow &  < \ln \; u_{i} > = \sum_{i=1}^{W} p_{i} \; \ln \; u_{i} \;\;\; ,
                      \label{eq:helmholtz} \\
E/k_{B}T & \rightarrow & - < \ln \; w_{i} > = - \sum_{i=1}^{W} p_{i} \; \ln \;
               w_{i} \;\;\; ,  \label{eq:energy} \\
 - \; \frac{\partial}{\partial x} (S/k_{B}) & \rightarrow & 
   \sum_{i=1}^{W} (\frac{\partial p_{i}}{\partial x}) \; \ln \; p_{i} \;\;\; ,
               \label{eq:canessa} 
\end{eqnarray}
with $k_{B}$ a constant added to have reduced units.  The choice of
the sign for $E$ is to approximate the system energy as the
sum of individual contributions (or mean energy)
$E \rightarrow <E> = \sum_{i} p_{i} \epsilon_{i}$, where 
$\epsilon_{i}/k_{B}T = - \ln \; w_{i}$ represents their energy states, and thus
obtain
\begin{equation}\label{eq:wi}
w_{i} = exp ( -\epsilon_{i}/k_{B}T )  \;\; \ge 0 \;\;\; .
\end{equation}
This relation is also found to be the solution of the differential equation:
$\frac{\partial \ln \; (p_{i}/u_{i})}{\partial (\epsilon_{i}/k_{B}T)}=-1$.

Furthermore from Eqs.(\ref{eq:sumderp}) and (\ref{eq:canessa}) it follows
immediately that the entropy satisfies
\begin{equation}\label{eq:entropy}
 \frac{\partial}{\partial x} (S/k_{B}) =
 - \; \frac{\partial}{\partial x} \left( \; \sum_{i=1}^{W} p_{i} \; \ln \; p_{i} \; \right) \;\;\; ,
\end{equation}
which is the basic result stated by the BG entropy of Eq.(\ref{eq:bg}).

Since the moving particles are assumed to form an elastic body, then the 
total work along an infinitesimal elongation $d\ell$ is given
by $d{\cal W} =dA = f_{\ell}\; d\ell$ for a change at constant 
$T$, with $f$ a ({\it compressive} $<0$, or
{\it expansive} $>0$) tensile force in the direction of $\ell$.  This relation
is obtainable directly from the first and second laws of thermodynamics in
a reversible isothermal process \cite{Tre75}, assuming that there is not 
another force acting on the system in addition to $f_{\ell}$ as for example
due to hydrostatic pressure $P$ ($d{\cal W}=-PdV \approx 0$; recall
volume $V$ here is an independent variable).
Therefore, the dynamics of our particle system relating thermodynamic
quantities reduces to
\begin{equation}\label{eq:newton}
F =  \left(\; \frac{\partial A}{\partial x} \;\right)_{T} = 
k_{B}T \; \frac{\partial}{\partial x} \left( \; \sum_{i=1}^{W} p_{i} \; \ln \; u_{i} \; \right)_{T} =
 \left(\; \frac{\partial E}{\partial x} \;\right)_{T} - \; 
 T \left(\; \frac{\partial S}{\partial x} \;\right)_{T} \;\;\; .
\end{equation}
Thus through Eqs.(\ref{eq:helmholtz}) and (\ref{eq:energy})
the sytem entropy can be correlated to Newton's law of motion in relation
to a distinct tensile force acting on the system.

\section{Remarks}

At a first glance the choice for $p_{i}$ in Eq.(\ref{eq:probability}) 
may be seen quite arbitrary, but in view of the results derived for $S$, 
$A$ and $E$ starting only from this quantity it seems to be reasonable. 
As shown below, this choice can further be justified in relation with the 
measure for an inverse temperature for the system.  
Using Eqs.(\ref{eq:probability}),
(\ref{eq:energy}), (\ref{eq:wi}) and (\ref{eq:entropy}), we obtain at once
\begin{equation}\label{eq:temperature}
\frac{\Delta (S/k_{B})}{\Delta (E/k_{B}T)} = 1 +
    \frac{\ln \; u_{_{W}}}{\ln \; w_{_{W}} } =
   1  - \left( \frac{k_{B}T}{\epsilon_{_{W}}} \right) \ln \; u_{_{W}} \;\;\; ,
\end{equation}
where $\Delta$ represents functions difference between consecutive
microstates fluctuations.  

Although we do not know how each of our $u_{i}$ and $w_{i}$ functions
may depend explicitly on the displacement $x$, we can still estimate all 
thermodynamic relations from given average values of either $u$ or $w$ 
at each state $i$.
For example, interesting to note are the results obtained when setting
$u_{i}=1/Z$, a sort of mean value for $u$ ($\forall i$) and $w_{i}$
variable.  From Eqs.(\ref{eq:probability}) and (\ref{eq:wi}), it follows 
then that $p_{i} = \frac{e^{-\epsilon_{i}/k_{B}T}}{Z}$, which
corresponds to the Gibbs distribution in thermal equilibrium
and satisfies $\frac{dp_{i}}{d\epsilon_{i}} = - \frac{p_{i}}{k_{B}T}$.
Using the normalization condition for $p_{i}$ we derive in turn the 
partition function $Z = \sum_{i} e^{-\epsilon_{i}/k_{B}T}$, where the
summation is over all microstates of the system.  From this
result and that of the energy states $\epsilon_{i}$ plus Eq.(\ref{eq:energy}),
we obtain a relation for the mean energy of a system at temperature $T$ as
$<E>=-\left( \frac{\partial \ln \; Z}{\partial (1/k_{B}T)}\right)$.
For the Helmholtz free energy of Eq.(\ref{eq:helmholtz}) we can write
$A/k_{B}T  = - \ln \; Z$, which is also a well-known classical statistical
mechanics result \cite{Man88}.

Let us analyse next a second case, {\it i.e.}, the case of having an averaged
$w$ ($\forall i$) and $u_{i}$ as the variable.  To derive Tsallis 
statistics we assume the thermal energy of the particles $k_{B}T$ 
to be proportional to their energy states $\epsilon$ by the nonextensivity 
(integer) factor $q-1$ for all the $i$-microstates.  From Eq.(\ref{eq:wi}) 
and $q \ne 1$ this is equivalent to consider
\begin{equation}\label{eq:wiq}
\ln \; w_{i} =  \frac{1}{1-q}   \;\;\; .
\end{equation}
Using the $\ln_{q}$ definition of Eq.(\ref{eq:lnq}) in conjunction with 
Eqs.(\ref{eq:probability}) and (\ref{eq:wiq}), we then get the relation
$- \ln_{q} \; (1/p_{i}) - \ln \; (p_{i}/u_{i}) = \frac{p_{i}^{q-1}}{q-1}$.
For $q \rightarrow 1$ and by Taylor expansion \cite{Tsa03} we can also approximate 
$p_{i}^{q} = p_{i} p_{i}^{-(1-q)} = p_{i} exp ( \ln \; p_{i}^{-(1-q)}) \approx p_{i} 
[ 1 + (1-q) \ln \; p_{i}^{-1} ]$. Thus in this limit 
$\ln_{q} \; p_{i}^{-1} \approx \ln \; p_{i}^{-1}$. 
These results imply that
\begin{equation}\label{eq:uiq}
\ln \; u_{i} \approx  \frac{p_{i}^{q-1}}{q-1}   \;\;\; .
\end{equation}
Therefore within our statistical thermodynamic approach it follows that
\begin{equation}
A/k_{B}T \approx \sum_{i=1}^{W} p_{i} \left( \frac{p_{i}^{q-1}}{q-1} \right) \;\;\; , \;\;\;
E/k_{B}T \approx - \sum_{i=1}^{W} p_{i} \left( \frac{1}{1-q} \right) = 
                \epsilon/k_{B}T  \;\;\; , 
\end{equation}
whence
\begin{equation}
S/k_{B}  = E/k_{B}T - A/k_{B}T \approx \frac{1 - \sum_{i=1}^{W} p_{i}^{q}}{q-1} \;\;\; ,
\end{equation}
which corresponds to the entropy term of Eq.(\ref{eq:tsallis}) introduced by Tsallis.
As in Ref.\cite{Tsa03}, if we assume equiprobability $p_{i}\equiv 1/W$ ($\forall i$) 
we also obtain $S=\ln_{q} \; W$.

Possible values of the $q$ index 
are constrained by the logarithm functions of Eqs.(\ref{eq:wiq}) and (\ref{eq:uiq}). 
The condition $0 \le p_{i} \equiv u_{i}w_{i} \le 1 \; (\forall i)$ implies that
$0 \le exp(1/(1-q)) \; exp(p_{i}^{q-1}/(q-1)) \le 1$.  Hence we can deduce that
$\Theta_{i}(q)\equiv (p_{i}^{q-1}-1)/(q-1) \le 0$, due to the positive and continue
nature of the exponential function.  Then the limiting behaviour of this function as 
$q \rightarrow 1$ is obtained by L'H\^opital's rule because this is an indeterminate
form of type $\frac{\infty}{\infty}$.  We found 
$\lim _{q \rightarrow 1}\Theta_{i}(q) = \ln p_{i}$, hence $0 \le exp( \; \ln p_{i} \;) \le 1$.  
From this result we also have 
$0 \le p_{i}^{q-1} \le 1 \; (\forall i, \forall q)$,
which via Eq.(\ref{eq:wiq}) it implies that $\ln w_{i} \le \Theta_{i}(q) \le 0$.  
This, in turn, implies a positive $w_{i}$ in agreement with Eq.(\ref{eq:wi}) such 
that $0 < w_{i} \le 1$.  Therefore we can conclude that the range of validity of 
our statistical thermodynamic model for Tsallis statistics is 
$q \; \varepsilon \; (\pm \infty,1]$. 
 
On the other hand, using Eqs.(\ref{eq:wiq}) and (\ref{eq:uiq})
for ($u$,$w$) at the upper microstate $i = W$, Eq.(\ref{eq:temperature}) 
for $q \ne 1$ takes the form
\begin{equation}\label{eq:temperature1}
\frac{\Delta (S/k_{B})}{\Delta (E/k_{B}T)} \approx 1 - p_{_{W}}^{q-1} \;\;\; .
\end{equation}
This implies that the thermodynamic definition for the temperature
at constant volume and number of particles 
$T=\left( \frac{\partial E}{\partial S}\right)_{V,N}$ follows for
$q >>1$ since, in physical terms, $p_{_{W}} \le 1$ for large {\it subextensive}
systems.  For {\it superextensive} systems with integers $q < 1$,
this relation becomes negative.  For $q=1$, from Eq.(\ref{eq:temperature})
we readily found $\frac{\Delta (S/k_{B})}{\Delta (E/k_{B}T)} = 1$ as should be
expected since $S_{q=1}$ recovers $S_{BG}$.

Finally for the force on the system within Tsallis statistics we have 
\begin{equation}
F = k_{B}T \; \frac{\partial}{\partial x}
     \left( \; \frac{\sum_{i=1}^{W} p_{i}^{q}}{q-1} \; \right)_{T} = ma \;\;\; , 
\end{equation}
which gives a physical significance to the $q$-values representing complex systems
and correlates them to the system dynamics.

In summary, the new insights gained by our statistical thermodynamic
approach of moving particles forming an elastic body are on the molecular-mechanical
properties derived for classical and nonextensive dynamical systems 
including a possible correlation between force and entropy in the system.
Our initial factorization for $p_{i}=u_{i}w_{i}$ may still be seen as rather arbitrary, 
but we have shown that this simple multiplicative form leads to interesting connections 
between an applied stretching force (tension) and thermodynamics quantities
of dynamical systems \cite{Can03}.  Such class of normalized product of positive
functions appears formally, {\it e.g.} in the analysis of stochastic processes
on graphs according to the Hammersley-Clifford Theorem.

\end{document}